\documentclass[fleqn,usenatbib,useAMS]{mnras}

\usepackage{graphicx}	% Including figure files
\usepackage{amsmath}	% Advanced maths commands
\usepackage{amssymb}	% Extra maths symbols
\usepackage{multicol}        % Multi-column entries in tables
\usepackage{bm}		% Bold maths symbols, including upright Greek
\usepackage{pdflscape}	% Landscape pages
\usepackage{journals}

\usepackage{etoolbox}
\makeatletter
\newcount\c@additionalboxlevel
\newcount\c@maxboxlevel
\AtBeginShipout{%
    \ifnum\value{additionalboxlevel}>\value{maxboxlevel}%
    \typeout{Warning: maxboxlevel might be too small, increase to %
        \the\value{additionalboxlevel}%
    }%
    \fi
    \@whilenum\value{additionalboxlevel}<\value{maxboxlevel}\do{%
        \typeout{* Additional boxing of page `\thepage'}%
        \setbox\AtBeginShipoutBox=\hbox{\copy\AtBeginShipoutBox}%
        \stepcounter{additionalboxlevel}%
    }%
    \setcounter{additionalboxlevel}{0}%
}
\makeatother

\def\HI{H~{\sc i}\, }
\def\HII{H~{\sc ii}\, }

\def\kms{$\textrm{km~s$^{-1}$}$}
\def\nb{\textsc{nbursts}}

\usepackage[T1]{fontenc}
\usepackage{ae,aecompl}

\usepackage{newtxtext,newtxmath}
\title[]{NGC~90: a hidden jelly-fish galaxy?}
\author[Zasov et al.]{
Anatoly V. Zasov,$^{1,2}$\thanks{E-mail:zasov@sai.msu.ru}
Anna S. Saburova,$^{1,3}$
Oleg V. Egorov,$^{1}$
Alexey V. Moiseev $^{1,4}$
\\
$^1$ Sternberg Astronomical Institute, Moscow M.V. Lomonosov State University, Universitetskij pr., 13,  Moscow, 119234, Russia\\
$^2$ Faculty of Physics, Moscow M.V. Lomonosov State University, Leninskie gory 1,  Moscow, 119991, Russia \\
$^3$ Institute of Astronomy, Russian Academy of Sciences, Pyatnitskaya st., 48, 119017 Moscow, Russia\\
$^4$ Special Astrophysical Observatory, Russian Academy of Sciences, Nizhniy Arkhyz, Karachai-Cherkessian Republic, 357147, Russia\\
}

\begin{document}
\large%\hfill{``cobs-6.tex'' РїС—Р…РїС—Р… 17.04.0

\label{firstpage}
\pagerange{\pageref{firstpage}--\pageref{lastpage}} \pubyear{2018}
\maketitle

\begin{abstract}
We study  a peculiar galaxy NGC~90, a pair member of interacting system Arp~65 (NGC~90/93), using the long-slit spectral observations carried out at the Russian 6m telescope BTA  and the available SDSS photometric data. This galaxy demonstrates two tidal tails containing young stellar population, being an extension of its `Grand Design' spiral arms. We obtained the  distribution of velocity and oxygen abundance of emission gas (O/H) for two slit orientations. In the central part of the galaxy a significant role belongs to non-photoionization mechanism of line emission probably caused by shocks due to LINER-like activity of the nucleus. The O/H  has a shallow abundance gradient, typical for interacting galaxies. The most intriguing peculiarity of the galaxy is the presence of the discovered earlier  huge \HI `cloud' containing about half of total mass of galaxy gas, which is strongly displaced outwards and has a velocity exceeding at about 340 \kms ~the  central velocity of  the main galaxy. We found  traces of current star formation in the `cloud', even though the cloud is apparently not gravitationally bound with the galaxy. A possible   nature of the `cloud' is discussed. We argue that it presents  a flow of gas sweeping by ram pressure and elongated along  a line of sight. 
\end{abstract}
\begin{keywords}
galaxies: kinematics and dynamics,
galaxies: evolution
\end{keywords}

\section{Introduction}

NGC~90 = UGC~208  is a spiral galaxy of SABc type, a member  of the interacting pair  of galaxies NGC~90/93, known as the system Arp 65. The distance between two galaxies is about 3 arcmin (64 kpc) in projection which is about 3 optical diameter $D_{25}$ of NGC~90 (according to Hyperleda\footnote{\url{http://leda.univ-lyon1.fr/} \citep{Makarov2014}}). This galaxy possesses a pair of regular spiral arms of Grand Design type, which are unusually contrast even at the NIR images. { Such feature of the spiral structure can be considered as a consequence of the interaction }\citep{Casteels2013, Oh2008}. Both arms straighten at the periphery, passing into the  tidal tails.  A thin and long  tail is aimed to north-west, and smooth and  a shorter one -- in the opposite direction.  Curiously, the line-of-sight (LOS) velocities of paired galaxies differ by more than 300 \kms (HyperLeda), and more than 400 \kms according to \HI velocity found by  \citet{Sengupta2015}.  Such velocity discrepancy is unusually high for  interacting galaxies with tidal structures. Both galaxies are the members of a group SRGb063,  where X-ray gas was detected \citep{Mahdavi2004}. 

Following \citet{Sengupta2015}, we assume that the distance to this system is 73 Mpc. Total K-band magnitudes of galaxies taken from Hyperleda correspond to the NIR luminosities $\log L_K$ = 10.69 (in  solar units) and 11.19 for NGC~90 and NGC~93 respectively, so one can expect that NGC~90 is slightly less massive than its neighbour.

Single dish observations of \HI reveal the unusually wide double-hump profile of hydrogen line \citep{Springob2005}.  \citet{Sengupta2015} carried out the resolved observations of NGC~90 and its tidal debris at  21 cm line.  They found  that a total mass of \HI connected with this galaxy is quite normal (about $7\times10^9 M_\odot$). However, about a half of hydrogen mass looks strongly displaced with respect to the main body -- both in space  and in the velocity field, which accounts for a wide profile of non-resolved observations. It looks like a  huge \HI `cloud' comparable with the optical galaxy by its size, which is projected on the  SE part of the disc. The LOS velocity of this \HI `cloud' does not agree with the expected velocity of a galaxy, exceeding at about 340 \kms ~the  velocity of galaxy centre. A cause of such peculiarities is not fully understood \citep[see discussion][]{Sengupta2015}, which makes this galaxy  very interesting for further investigation.

 \begin{figure*}
\centerline{
 \includegraphics[height=0.47\linewidth]{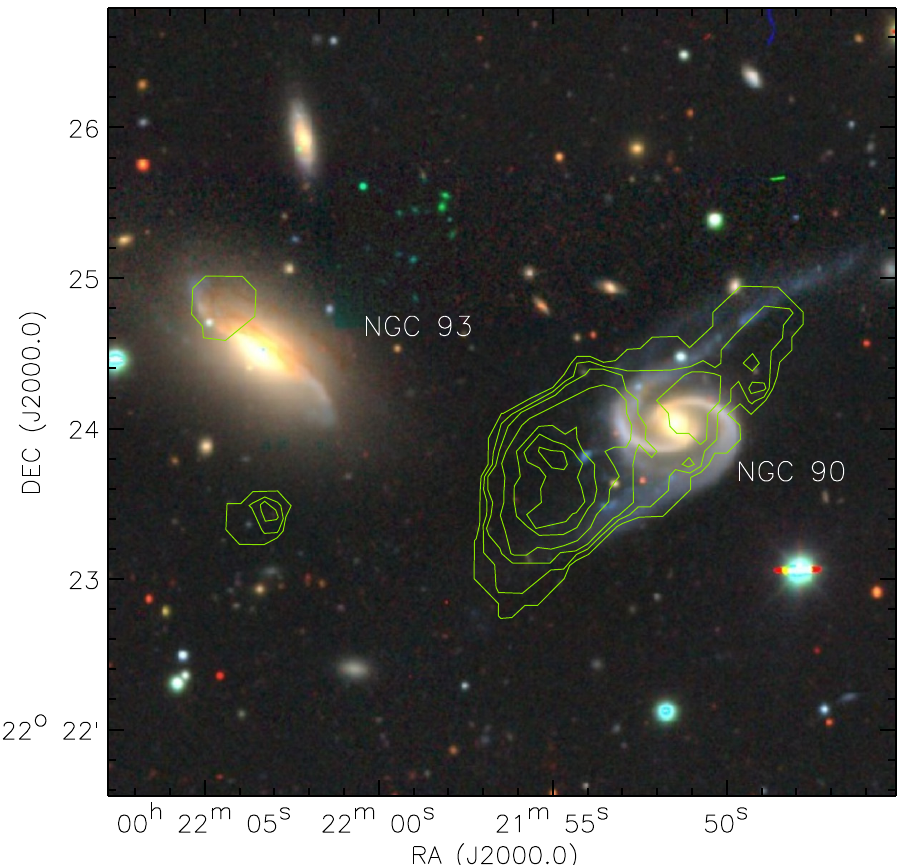}
 \includegraphics[height=0.47\linewidth]{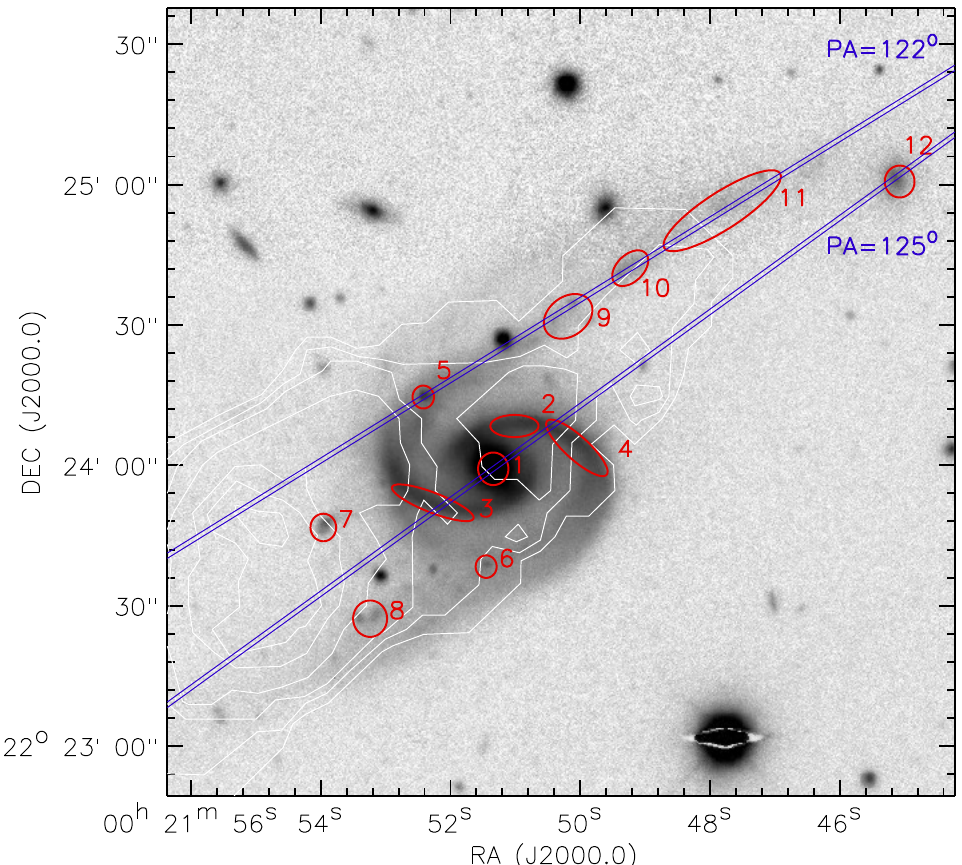}
 }
\caption{Right: composite {\it g,r,i}-band DECaLS DR8 image of Arp 65 pair  with contours of \HI surface density from \citet{Sengupta2015}. Left: the zoomed region around NGC~90 in DECaLS  {\it r}-filter with \HI contours 
and  overplotted  positions of the slit and the apertures considered in photometric analysis.}\label{map}
 \end{figure*}

In this paper which is a continuation of our series of works on the interacting systems \citep{zasovetal2015, zasovetal2016,Zasovetal2017, Zasovetal2018, Zasov2019} we describe the results of optical spectral observations of  NGC~90 with a long-slit  aimed to study its kinematics and gas-phase abundance  and to clarify a possible nature of peculiarities of this galaxy.  DECaLS DR8 image of the system is given in Fig.\ref{map} where slit positions and \HI contours from \citet{Sengupta2015}  are also shown. For several highlighted areas marked in the image of NGC~90, we used photometric estimates (see below).

\section{Observations and data reduction and analysis }\label{obs}
We carried out long-slit spectral observations of NGC~90 with  the Russian 6-m telescope with SCORPIO-2 multi-mode reducer of the telescope prime focus \citep{AfanasievMoiseev2011}. The observations took place on 4-th and 5-th of November 2015 for PA=125$^\circ$ and PA=122$^\circ$ correspondingly under the seeing 1.9 arcsec. The exposure times are 4500 and 7200 seconds for  PA=125$^\circ$ and PA=122$^\circ$.  We utilized the grism VPHG1200@540, which covers the spectral range 3600--7070 \AA\, and has a dispersion of 0.87 \AA\ pixel$^{-1}$. The spectral resolution is $\approx 5.2$ \AA\, estimated as FWHM of night-sky emission lines. The scale along the slit was 0.36~arcsec pixel$^{-1}$,  the slit width was  1~arcsec. We demonstrate  the positions of the slit in Fig.~\ref{map}. 

We processed the data using our \textsc{idl}-based pipeline. The reduction was already described in the previous papers \citep[see, e.g.][]{zasovetal2015, zasovetal2016}. It includes a bias subtraction and truncation of overscan regions, flat-field correction, the wavelength calibration based on the spectrum of He-Ne-Ar  lamp, cosmic ray hit removal, combination of individual exposures, the night sky subtraction and flux calibration. 

After the reduction we fitted the processed spectra taking into account  the parameters of  instrumental profile of the spectrograph resulted from the fitting  of the twilight sky spectrum observed in the same  observation runs. We convolved the parameters of  instrumental profile with the high-resolution PEGASE.HR~\citep{LeBorgneetal2004} simple stellar population models (SSP) and fitted the reduced spectra of galaxies.  We performed it using the \nb{} full spectral fitting technique  \citep{Chilingarian2007a, Chilingarian2007b}, which  allows to fit the spectrum in a pixel space.  In this method the parameters of the stellar populations are derived by nonlinear minimization
of the quadratic difference chi-square between the observed and model spectra. We utilized the following parameters of SSP: age T~(Gyr) and metallicity [$Z/H$]~(dex) of stellar population. The line-of-sight velocity distribution (LOSVD) of stars was parameterized by Gauss-Hermite series (see \citealt{vanderMarel1993}).

The emission spectra were obtained by subtraction of model stellar spectra from the observed ones. After that we fitted the Gaussian profiles to emission lines to estimate the velocity and velocity
dispersion of ionized gas and the fluxes in the emission lines. All measured flux ratios were corrected for reddening based on derived Balmer decrement using the reddening curve from \cite{Cardelli1989} parametrized by \cite{Fitzpatrick1999}.

Using the measured strong emission lines ratios we estimated the distribution of oxygen abundance (which is a good indicator of gas phase metallicity) along each slit.  Due to the low sensitivity of the CCD in the bluest part of the spectrum ($\lambda<4000$\AA), we were unable to measure flux of [O~\textsc{ii}]3727\AA\, line for most of the observed regions and hence we were very limited in a number of available methods that we can use to derive oxygen abundance. In this paper we rely our measurements on two empirical calibrations: O3N2 by \citet{Marino2013}, which utilize [O~\textsc{iii}]5007\AA/H$\beta$ and [N~\textsc{ii}]6583\AA/H$\alpha$ flux ratios, and S calibration by \citet{Pilyugin16}, which uses also [S~\textsc{ii}]6717+6731\AA/H$\alpha$ in addition to the mentioned flux ratios. Note that in the original version of the S method all the measured fluxes are normalized to H$\beta$, while here we use the reddening-corrected fluxes relative to the closest Balmer line (H$\alpha$ or H$\beta$) and rely on the theoretical flux ratio of H$\alpha$/H$\beta$ = 2.86 as expected for \HII regions at the electron temperature $T_e=10000$K \citep{Osterbrock2006}.

 \subsection{Results of spectral observations}
 
\begin{figure*}
\includegraphics[width=0.5\linewidth]{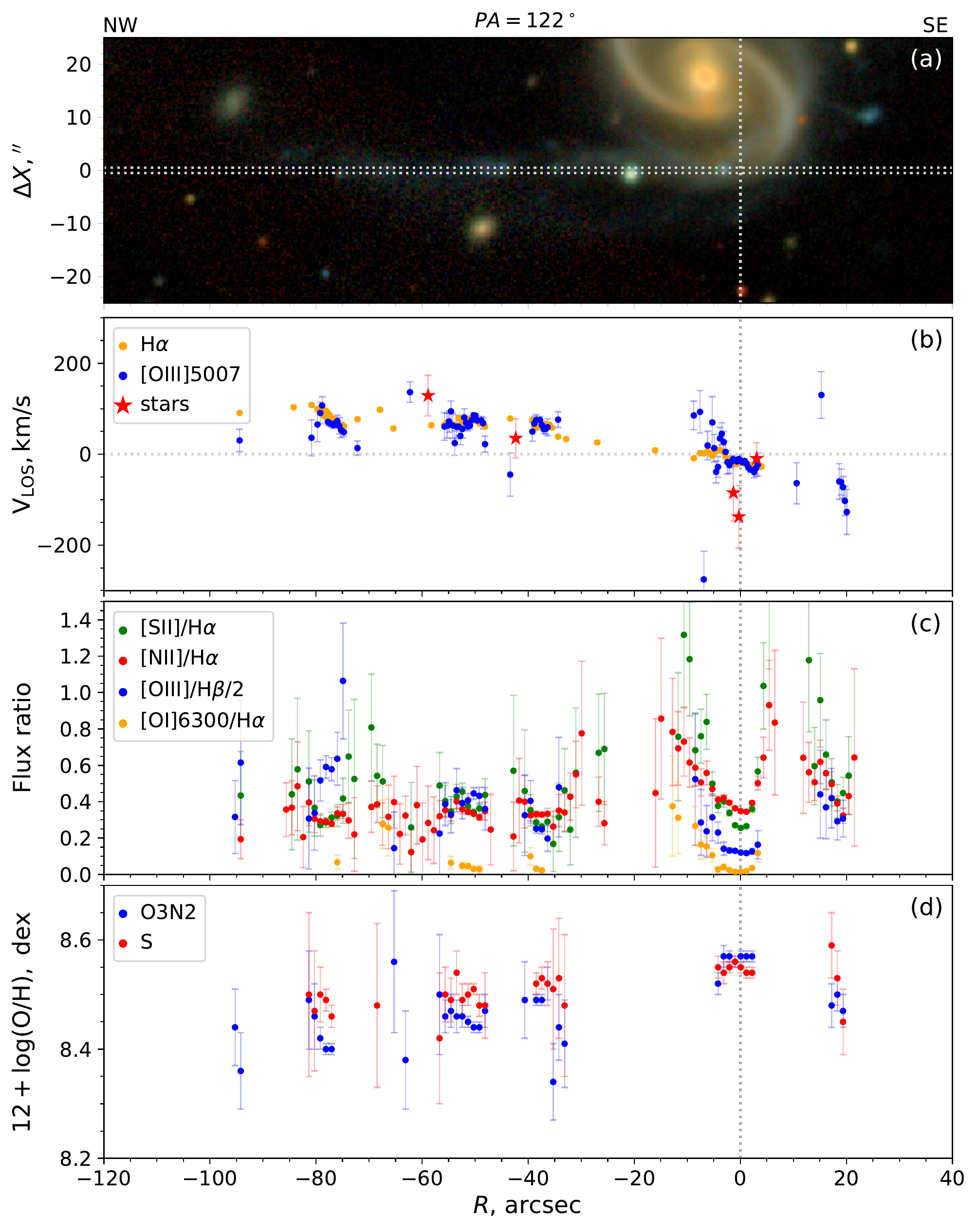}%14.5cm
\includegraphics[width=0.5\linewidth]{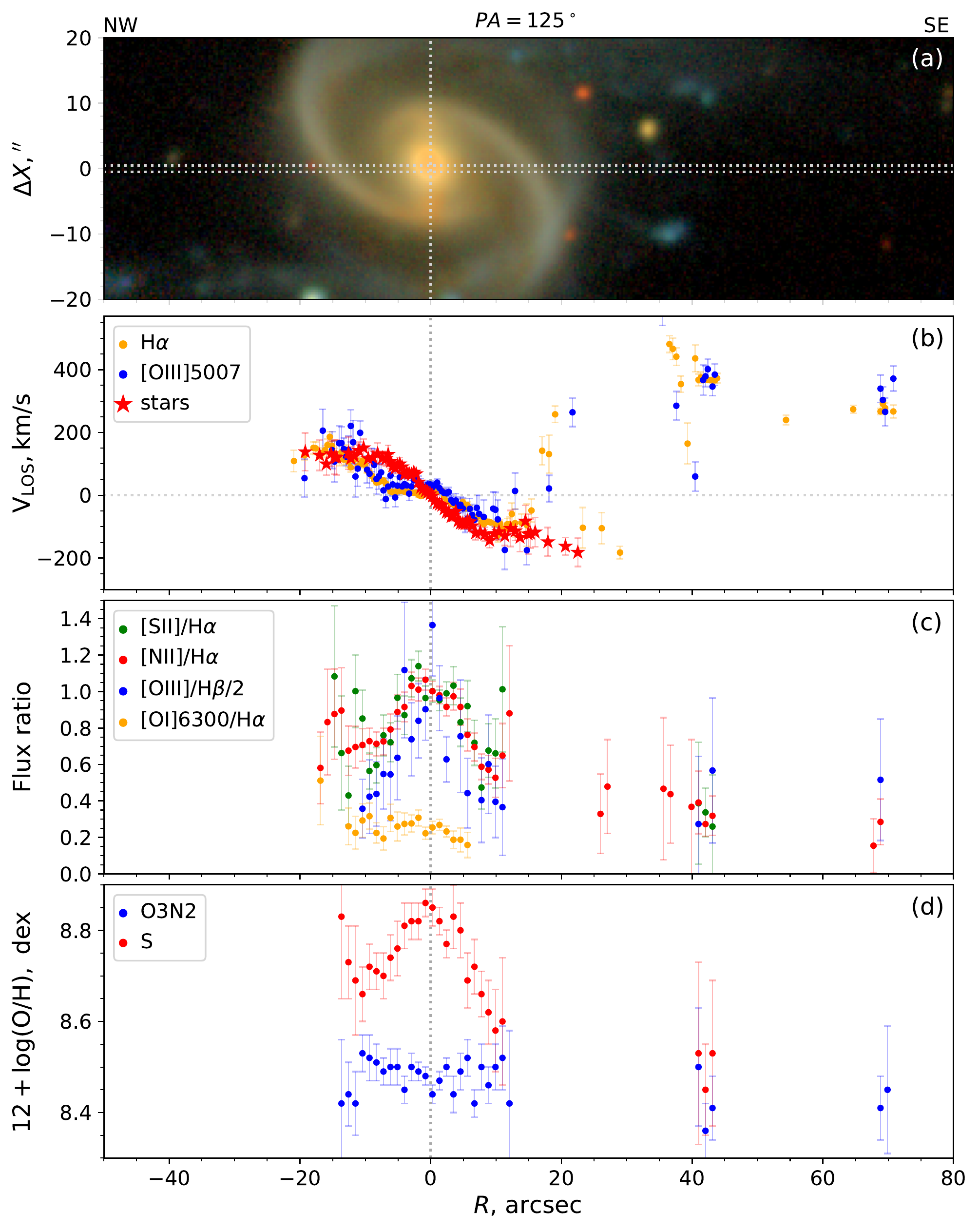}%14.5cm
\caption{Radial variation of the parameters estimated from the fitting of the spectra of NGC~90 for PA=122$^\circ$ (left) and  PA=125$^\circ$ (right). From top to bottom we give: the composite DECaLS {\it g,r,z}- band image with overlaid position of the slit {(the slit width is shown as the distance between two horizontal lines)}; the radial variation of: (b) LOS velocities relative to 5197 \kms\, -- velocity of the NGC~90 optical centre; (c) strong emission lines flux ratios,  and (d) oxygen abundance derived by O3N2 and S empirical methods \citep[][respectively]{Marino2013,Pilyugin16}. Note that all points corresponding to the non-photoinisation mechanism of excitation according to Fig.~\ref{arp65_bpt} or having $\mathrm{EW(H\alpha)<3}$~\AA\, were removed from panel (d) for PA=122$^\circ$ while no any selection were made for PA=$125^\circ$ (see text for details).}
\label{arp65_results}
\end{figure*}

The results of the analysis of our spectral data are given in Fig. \ref{arp65_results}, which demonstrates the distribution of measured parameters along the slits with PA=122$^\circ$ (left) and PA=125$^\circ$ (right).
For each slit we show: (a) position of the slit marked by horizontal dotted line overlaid on the DECaLS {\it g,r,z}-band image, (b) LOS velocity relative to that of the optical centre of NGC~90 \citep[5197 \kms\, according to][]{Sengupta2015}, (c) strong emission lines flux ratios\footnote{Note that flux ratios of [O~\textsc{iii}]/H$\beta$ and corresponding uncertainties were divided by 2 to cover the same range as for other lines on the plot}, and (d) oxygen abundance derived by O3N2 and S methods (see Section~\ref{obs} for details). 

\begin{figure*}
\centerline{\includegraphics[width= \linewidth]{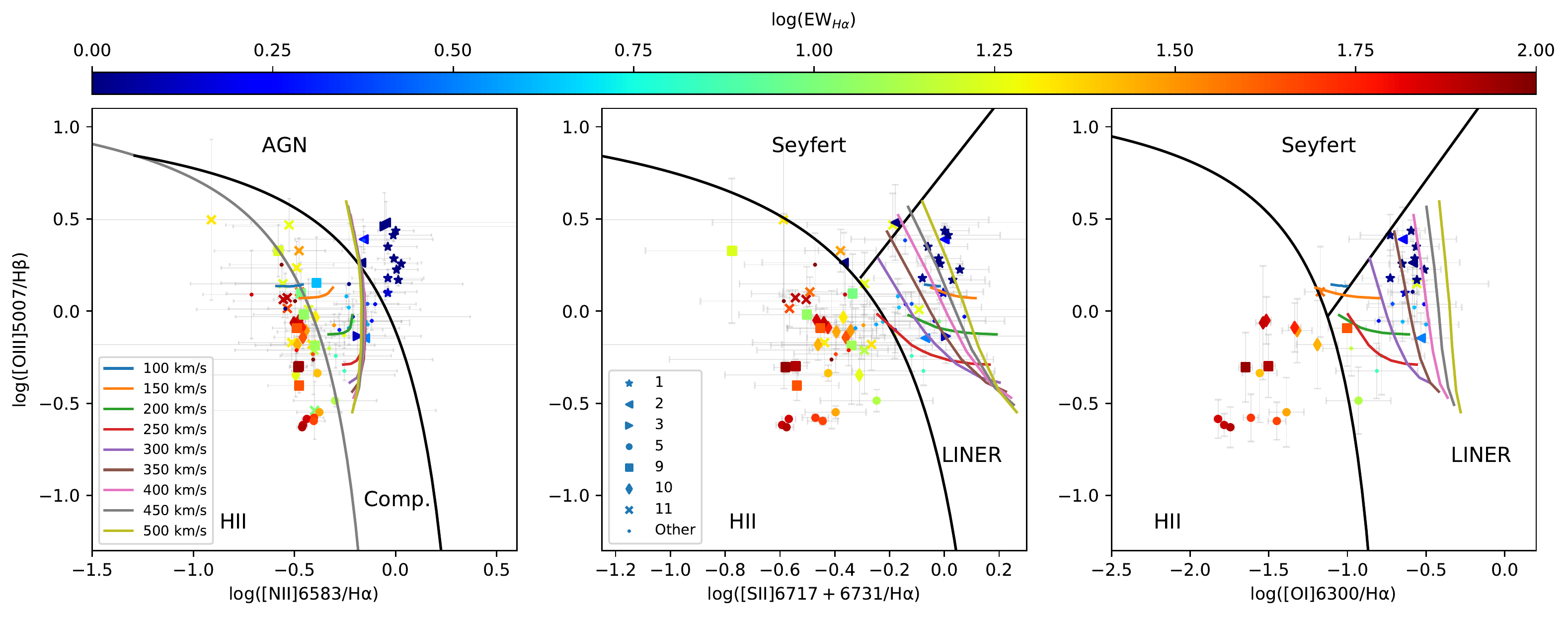}}%14.5cm
\caption{Diagnostic diagrams plotted for the regions for which we  obtained spectral data. The logarithm of  EW(H$\alpha$) is colour-coded. Different symbols correspond to the regions listed on the central panel according to their denotation in Fig.~\ref{map} and Table~\ref{colors}. Black curve is the `maximum starburst line' \citep{Kewley2001} separating theoretical photoionised \HII-regions and all other types of gas excitation. Grey line separates the regions with composite mechanism of excitation according to  \citet{Kauffmann03}. Black diagonal line from \citet{Kewley2006} separates the areas occupied by LINER and Seyfert on the diagrams. Colour curves are the lines of a constant velocity of shocks according to the model from \citet{Allen2008} for \citealt{Dopita2005} chemical abundance and without precursor.}
\label{arp65_bpt}
\end{figure*}

The panel (b) of Fig.~\ref{arp65_results} for the slit with PA=$122^\circ$ demonstrates how the LOS velocity changes along the northern tidal tail. The gas velocity slowly grows outwards and flattens at large distances. Velocity of the outer tail regions beyond the main body of NGC~90 differs by about 100 \kms\, from that of the galactic centre (V=0 in our diagram). 
{ One can see} that both gaseous and stellar components demonstrate a similar behaviour, hence the  stars and the emission gas in these areas move together inseparably. 
 
A LOS velocity distribution along the second slit (PA=$125^\circ$) running through the center of galaxy at an angle not far from the dynamic major axis (see below) show a well ordered integral-shaped pattern of rotation of stellar disc with the semi-amplitude $\sim$ 130 \kms. This is in a good agreement with the rotation velocity $\sim$ 100 \kms\, derived from \HI data by \citet{Sengupta2015}. Unlike stars, ionized gas kinematics demonstrates a strong non-circular motion in the inner $\sim20$~arcsec: the difference between stars and gas exceeds 100 \kms\ NW of the nucleus both on forbidden and Balmer emission lines. This velocity difference is too high and asymmetric to be explained by shocks generated of the central small bar.  Most likely this kinematic decoupling feature is related with off-plane gas motions driven by  interaction  with NGC~93. With current spectral resolution we were able to reliably measure LOS velocity dispersion only in this central part of NGC~90. A stellar velocity dispersion there is 120 -- 140 \kms, which is quite normal for galactic bulges of mild luminous galaxies. The velocity dispersion of ionized gas in $H\alpha$ line reaches 60--100 \kms\, (with maximum in the centre) confirming the non-circular motions of gas in the circumnuclear region.

The most unusual feature of the velocity distribution along the slit PA=$125^\circ$ is the extended regions of abnormally high velocity of gas at  $R\sim20$~arcsec  and  at $R\sim35--70$~arcsec (to the south-east from the galaxy), which are hardly visible in the optical image.  The LOS velocity of the farther areas exceeds the central velocity of the galaxy at about 300 \kms, and the difference is even higher with respect to the SE part of the disc, on which they are projected. Curiously, this high-velocity emission `island' is close both by LOS velocity and by its position to the extended H~\textsc{i} cloud found by \citet{Sengupta2015}, projected onto SE part of a galaxy (they called it `SE debris'). According to these authors, the LOS velocity of SE debris found from \HI data, is $5531\pm 27$ \kms,  which is also higher at 343 \kms\, than their estimate of \HI velocity of NGC~90.

A distribution of the strong emission lines flux ratios in panel (c) of Fig.~\ref{arp65_results} demonstrates a different ionization state of the gas in the tidal tail and within the central part of the galaxy. Thus, the ratios of [N~\textsc{ii}]6583\AA/H$\alpha$, [S~\textsc{ii}]6717+6731\AA/H$\alpha$ and [O~\textsc{iii}]5007\AA/H$\beta$ for the clumps in the tidal tail (and also for the faint region at $R\sim42$ arcsec along the slit PA=$125^\circ$) are typical for \HII regions, while a strong enhancement of these ratios is observed toward the centre of the galaxy, indicating that a dominant mechanism of excitation there differs from classical  photoionisation. In the tidal tail, the [N~\textsc{ii}]/H$\alpha$ and [S~\textsc{ii}]/H$\alpha$ grows outside from the star-forming clumps, and the ionization state between them is typical for diffuse ionized gas (DIG) \citep[see, e.g.][]{Haffner2009}.

In order to better analyse the ionization state of the gas in different regions, we plot the diagnostic diagrams in Fig.~\ref{arp65_bpt}, namely (from left to right), the BPT diagram \citep*{BPT} showing the dependence of $\log$([O~\textsc{iii}]5007/H$\beta$) vs $\log$([N~\textsc{ii}]6583/H$\alpha$), and its extended versions for $\log$([O~\textsc{iii}]5007/H$\beta$) vs $\log$([S~\textsc{ii}]6717+6731/H$\alpha$) and $\log$([O~\textsc{iii}]5007/H$\beta$) vs $\log$([O~\textsc{i}]6300/H$\alpha$) \citep{BPT_S}. We refer further  all of them as BPT-diagrams. In this Figure we plot all measurements along the slits that correspond to a given  selected area (according to denotation in Fig.~\ref{map}) by the same symbol, while the colour of the symbol is coded according to the measured EW(H$\alpha$) -- equivalent width of H$\alpha$ line (in logarithmic scale). To separate the areas of different mechanism of excitation we overlaid three demarcation lines on the BPT-diagrams. Theoretical `maximum starburst line' from \citet{Kewley2001} separates the regions that could be explained by pure photoionization (below the line) from those which demands additional source of ionization. Grey line from \citet{Kauffmann03} separates the regions of pure photoionization from those with composite mechanism of excitation. Third line from \citet{Kewley2006} separates the areas on BPT-diagrams typically populated by Seyferts and LINER (low-ionization nuclear emission-line region). 

As it follows from BPT-diagrams, the emission lines ratios demonstrate a clear bi-modality. The centre of NGC~90 and the spiral arms lie mostly above the `maximum starburst line' and have low EW(H$\alpha$)$<5$~\AA, while most of the regions in the tidal tail seem to be photoionized with few faintest points lying in the area of composite excitation. Following \citet{Lacerda2018}, we consider that the regions having EW(H$\alpha$)$<14$~\AA\, are highly contaminated by surrounding DIG, while those having EW(H$\alpha$)$<3$~\AA\ represent a pure DIG. According to this criteria, in the tidal tail the DIG contribution is small, being significant only in the outskirts of star-forming clumps, while the emission from spiral arms and from the galaxy centre is purely from DIG. 

The origin of DIG in galaxies is still under debate \cite[see, e.g.][for review]{Haffner2009, Zhang2017}. Taking into account the  enhanced velocity dispersion of ionized gas in the centre (together with its position on the BPT-diagrams), we suppose that shocks from the mentioned above non-circular gas motion or/and a LINER-like activity of nucleus are most probable explanation of the DIG emission there. Indeed, the models of shocks taken from \citet{Allen2008} could fit well the emission lines ratios of the central part of NGC~90 on BPT-diagrams. Colour curves in Fig.~\ref{arp65_bpt} correspond to the lines of a given shock velocity according to the grid of models computed for oxygen abundance $\mathrm{12+\log(O/H)=8.44}$ (more precisely -- the chemical abundance from \citealt{Dopita2005}) that is nearly equal to the measured for NGC~90 in this paper (see below). The lines ratios on BPT-diagrams could be reproduced by shocks with velocities of 300--450 \kms. The reason  for some disagreement between the observed [N~\textsc{ii}]/H$\alpha$ values and the model ones might be a slightly higher metallicity of the galaxy than  it was used in the models.
Thus, the models of shocks at nearly solar metallicity considered by  \citet{Allen2008} predict  significantly  higher values of [N~\textsc{ii}]/H$\alpha$ than those we found, however they agree with the other observed flux ratios.   Note also that the shocks models on BPT-diagrams correspond to pure shocks, without precursor. The models with precursor fail to explain the observed line ratios. In general,  we may conclude that the origin of the shocks in the centre of NGC~90 is not related to a star-forming activity. 

Panels (d) in Fig.~\ref{arp65_results} show a distribution of the oxygen abundance along the slits derived by two empirical methods. Before doing that, we masked all points lying above `maximum starburst line' on BPT-diagrams or having EW(H$\alpha$)$<6$~\AA. After applying that for the slit with PA=$125^\circ$, all points from the central part of the galaxy ($-15<R<15$~arcsec) were masked. However we decided to leave these points on that plot for further discussion (see below).

In general, the obtained value $\mathrm{12+\log(O/H)\,\approx8.5}$ for the galaxy nicely agrees with its absolute magnitude  $M_B = -20.2$    \citep[see, e.g.][]{Pilyugin2004} . Oxygen abundance derived for star-forming regions in the tidal tail (slit with PA=$122^\circ$) shows a mild gradient along the tail. Both O3N2 and S methods yield similar values of oxygen abundance within the uncertainties. In the brightest region (\#5 in Fig.~\ref{map}) $\mathrm{12+\log(O/H)=8.55\pm0.02}$, while it drops down to $\mathrm{12+\log(O/H)=8.45\pm0.08}$ in the outer part of tidal tail (region \#11). Thus, our measurements reveal a shallow gradient of (O/H) $\approx$ 0.1 dex within the radial distance interval of about 22 kpc\footnote{The approximate radial distance between regions \#5 and \#11 after correction for inclination adopting position angle of major axis $PA=120.1^\circ$ and inclination angle $i=34.5^\circ$ taken from HyperLeda}, or $\Delta\mathrm{(O/H)\sim0.005\ dex\ kpc^{-1}\approx 0.05\ dex\ R_{25}^{-1}}$\footnote{We assume $R_{25}=10.4$~kpc according to HyperLeda}, that is $6-8$ times lower than that for normal spiral galaxies \citep{Pilyugin2014, Ho2015}. This evidences a radial motion of gas inspired by interaction, which is typical for galaxies in close pairs \citep{Kewley2010, Rosa2014} and for interacting or merging galaxies \citep{Rupke2010, Rich2012}. This estimate is also consistent with our previous measurements for other interacting systems \citep[see, e.g.]{zasovetal2015}.
 
The oxygen abundance towards the \HI cloud to the south-east of the galaxy is the same as in the northern tidal tail: $\mathrm{12+\log(O/H)=8.42\pm0.06}$ for the faint star-forming regions observed there,  crossed by the slit with PA$=125^\circ$. The oxygen  estimates obtained for the centre of NGC~90 by two methods significantly differ. However, as we noticed above, the emission from the central zone  comes from DIG, not from star-forming regions, and hence the empirical calibrators of oxygen abundace that based on measurements for \HII regions are not reliable (especially those like S method, which utilize [S~\textsc{ii}]/H$\alpha$ flux ratio that is significantly enhanced in DIG). On the other hand, in recent work by \citet{Kumari2019}, the authors showed that O3N2 method works well for objects highly polluted by DIG. Taking into account that the value of central oxygen abundance $\mathrm{12+\log(O/H)=8.47\pm0.02}$ revealed by O3N2 method is consistent with estimates made for outer star-forming regions in the galaxy and with the metallicity for the shocks models used during analysis of BPT-diagrams, we may expect that our estimates are reliable\footnote{Note however that \citet{Kumari2019} considered mostly the DIG ionized by leaking quanta from \HII regions, while in our case DIG has most probably shock-induced origin, so the oxygen abundance estimates in the centre of NGC~90 still should be used with a caution.}. If we adopt this estimate of (O/H) in the centre, then two features in the oxygen abundance distribution might be highlighted. First is the flat distribution of abundance in the central part (actually, a slight reversed gradient is observed, but it lies within the uncertainties for individual points). Second -- the central oxygen abundance is slightly lower than in the brightest star-forming region \#5 in the basement of tidal tail. Both these features might result from the possible gas infall to the central part of the galaxy that flattens the metallicity distribution. Such infall might be responsible for the mentioned above non-circular motions in the centre of the galaxy and for the shocks.

\subsection{Photometry and mass of NGC~90}

\begin{figure}
  \vspace{-13.0cm}
\includegraphics[width=4.0\linewidth]{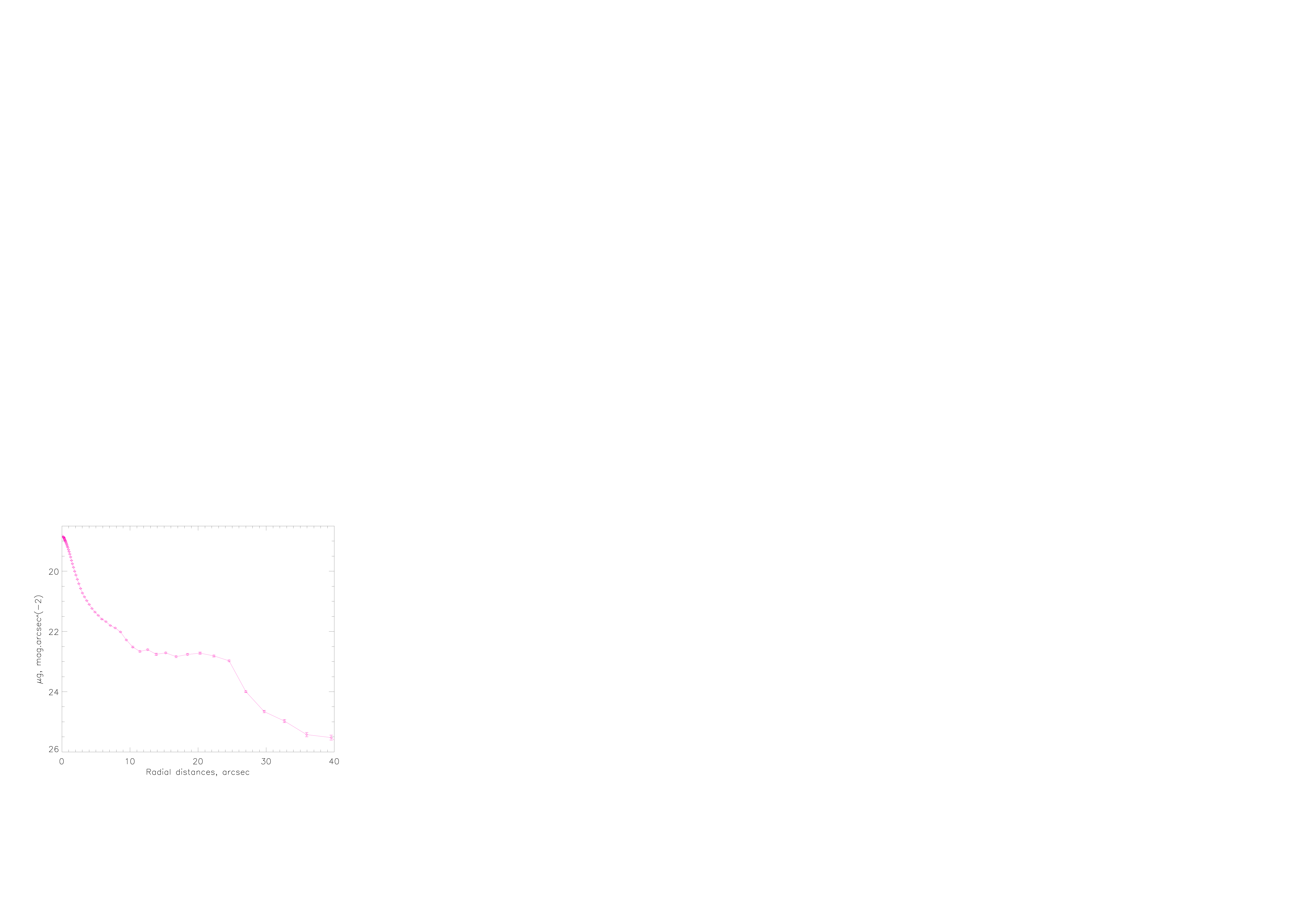}%14.5cm
\vspace{-4.0cm}
\caption{ The radial variation  of {\it g}-band surface brightness of NGC~90.  }
\label{arp65_mu}
\end{figure}

\begin{figure}
 \vspace{-13.0cm}
\includegraphics[width=4.0\linewidth]{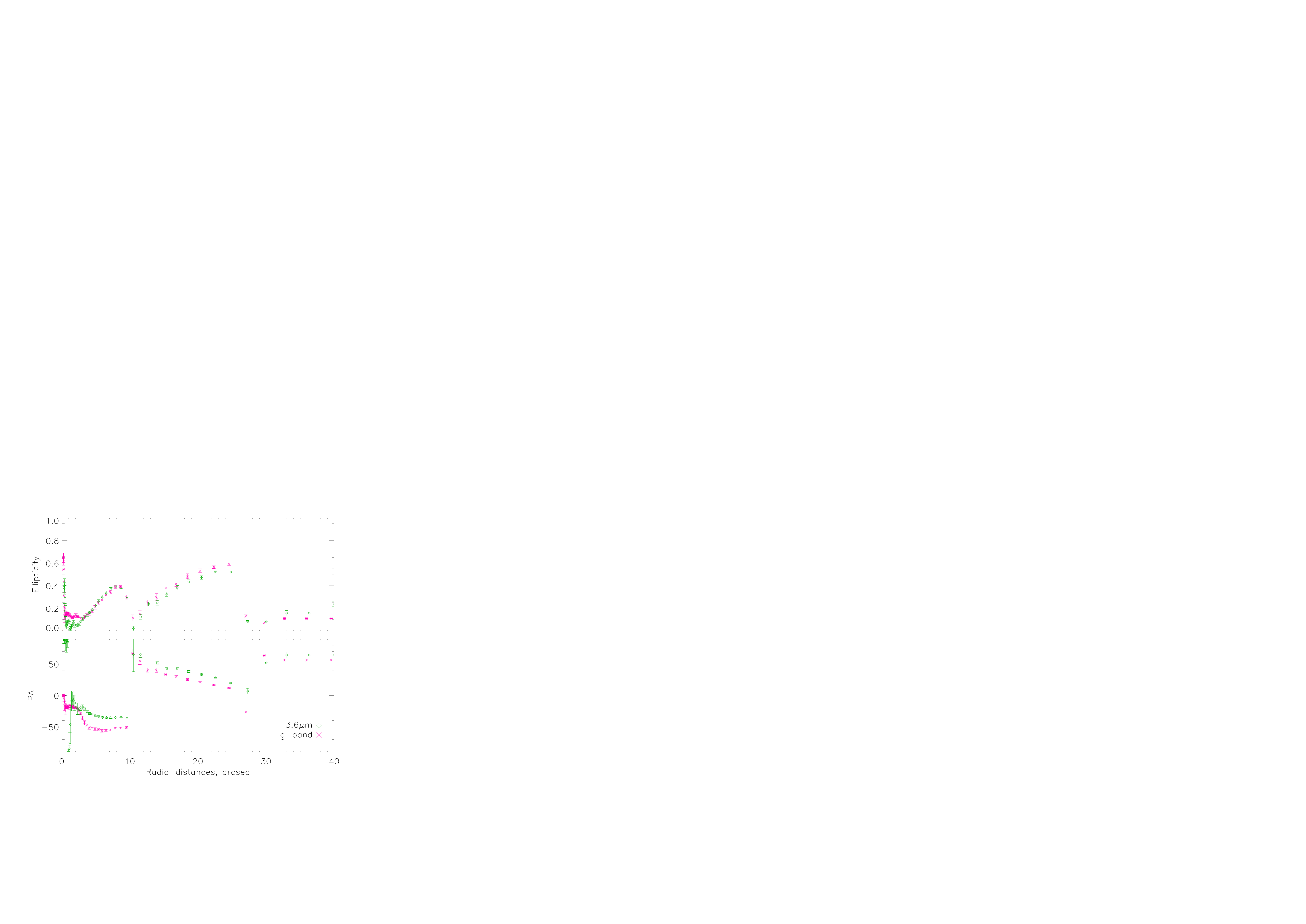}%14.5cm
\vspace{-4.0cm}
\caption{ Radial variation of the ellipticity (top panel) and position angle ({ bottom} panel) of NGC~90 for {\it g}-band (crosses) and 3.6$\mu$ m (diamonds).    }\label{arp65_ba}
\end{figure}

NGC~90 is a spiral galaxy with a very complex structure, as it is evident from the analysis of its photometric profiles. We analyzed the archive SDSS {\it g}-band and Spitzer IRAC 3.6 $\mu$m images of NGC~90 using the {\sc ellipse} routine \citep{Jedrzejewski1987} in the {\sc iraf} software environment (\citealt{iraf}).  Fig. \ref{arp65_mu} shows radial brightness distribution for {\it g}-band averaged within elliptical rings with variable ellipticity $\epsilon=1-b/a$ and position angle of major axes $(PA)_0$ of isophotes. In Fig. \ref{arp65_ba} we demonstrate  the variation of  $\epsilon$ and  ($PA)_0$ of isophotes with radial distance  in {\it g}-band and at 3.6 $\mu$m. Three zones can be distinguished on the diagrams: the central region within $R \sim 10$ arcsec, which includes bulge and a short bar,  the region of bright regular spiral arms, and the outer region, which, judging by its low ellipticity, we observe nearly face-on. 

Note that a radial variation of ellipticity or $(PA)_0$ along $R$ describes a brightness, not mass distribution. In particular, a change of $(PA)_0$ in the zone of spiral arms just reflects their curved shape.  Most probably, a tilt of a disc plane sharply changes at $R \approx 25''$ between the area of spiral arms and the outer low-inclined disc. At the same time,  non-asymmetric brightness distribution in the arms region prevents the assessment of real disc orientation there. 

%One way to find the most probable orientation of the inner disc under assumption that it is flat is to propose that the spiral arms in the main body of the galaxy become tightly wound before they transform  into straight tidal arms. Visually they fit nicely into the ellipse with  the ellipticity $\epsilon \approx 0.5$,  and $(PA)_0 \approx 100^\circ$ (see Fig. \ref{arp6536m}, where the  NIR image of galaxy is used). However this approach is rather illustrative, being not too reliable.

 The most certain way to specify $(PA)_0$  is to use a dynamic approach, namely to trace the LOS variation along the off-centre slit at PA = $122^\circ$.  If the motion of gas does not deviate strongly from the circular one,  one should expect that the zero LOS velocity V=0 with respect to the center of galaxy is observed where the slit cuts the dynamic minor axis.  Judging from the LOS velocity distribution, zero velocity falls on the interval $R=0-20$ arcsec.   Corresponding values of $(PA)_0 = 95-130^\circ$  may be considered as the most probable range of dynamic major axis of the inner disc. In this case we may identify a turnover of LOS velocity observed  at about 10 arcsec from the centre with the turning point  of rotation curve to a plateau at  $V_{los} \approx 130$ \kms,  as the velocity of disc rotation in projection onto the line-of-sight. After correction for projection for the adopted b/a=0.5, a circular velocity $V_{rot}$ beyond the turnover is $\sim$ 230 \kms\, for $(PA)_0 = 95^\circ$ and $\sim$150 \kms\, for  $(PA)_0 = 130^\circ$.  
 
 Note that the latter value of  $V_{rot}$ better agrees with Tully-Fisher relation.  Indeed, NIR magnitudes $m_H$ = 11.2  and $m_{K_s}$=11.0, according to the NED database for NGC~90, when applied to the Tully-Fisher relation  (see \citet{Ponomareva2017}), give us the expected `flat' velocity of rotation $\sim$ 170 \kms\, with the error about 0.1 dex. Rough estimations of total mass within the optical  diameter $D_{25}$= 21 kpc are  $\sim 13\times10^{10}M_\odot$ for  $(PA)_0 = 95^\circ$ and $\sim 6\times10^{10}M_\odot$ for  $(PA)_0 =130^\circ$. Thus, a dynamic  mass  within $D_{25}$ is determined at best with an accuracy of about  factor 2.
 
\begin{figure*}
\includegraphics[width=0.85\linewidth]{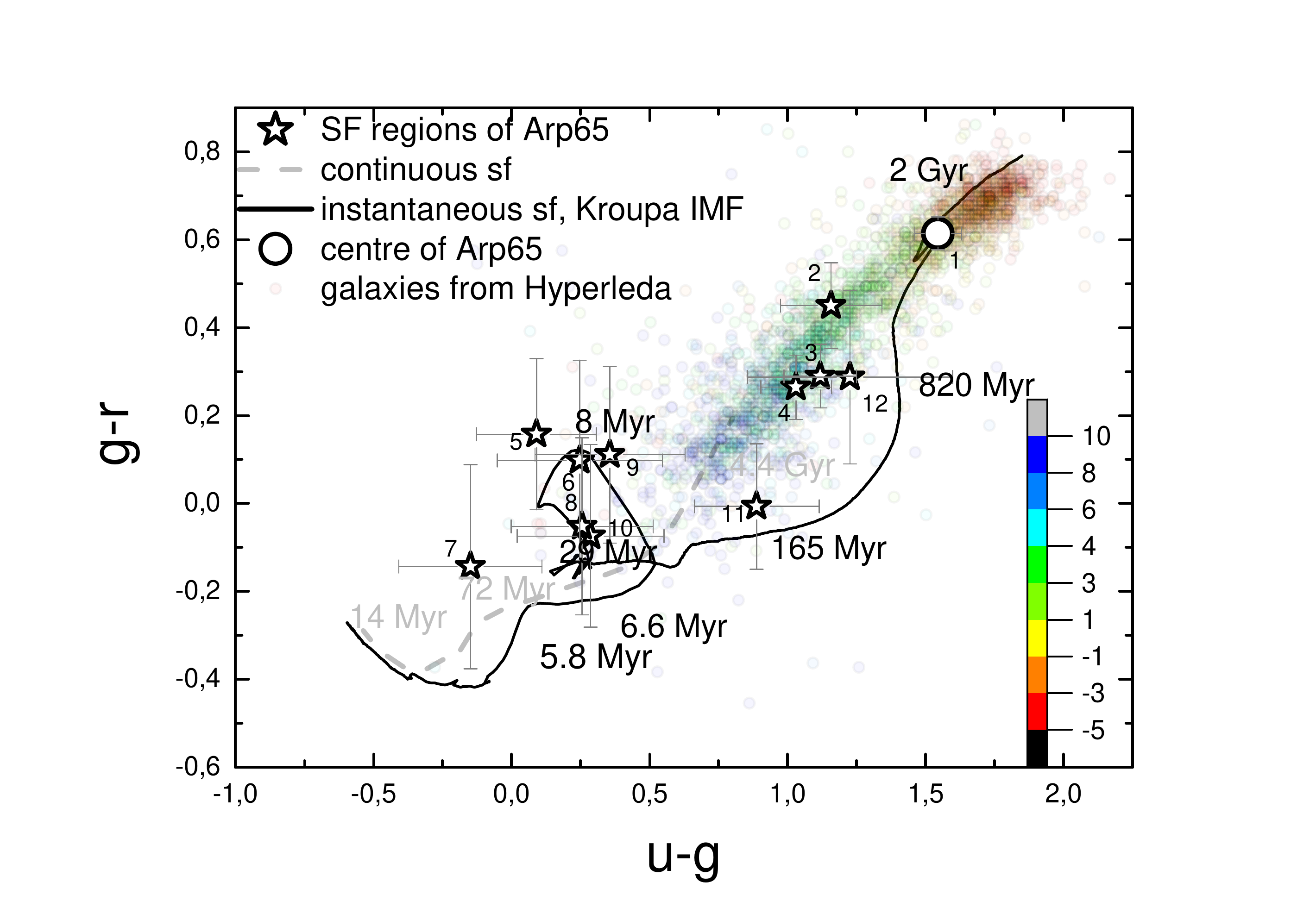}%14.5cm
\caption{ The position of the regions of NGC~90 shown in Fig.\ref{map} on the (g-r) versus (u-g) diagram (symbols with numbers). The position of centre of the galaxy  is given  by circle. Small transparent circles show for the comparison galaxies of various morphological types for which we got U-B and B-V colours from Hyperleda { database} and converted them to (g-r) and (u-g). The type is colour-{ coded}. The colour-bar is shown in the plot, where the numbers correspond to the types. Black line and gray dashed line with the signed ages correspond to Starburst99 model tracks for instantaneous and continuous star formation respectively. }\label{arp65_ug_gr}
\end{figure*}

  The  colour indices of several individual regions of enhanced brightness, obtained from SDSS images and marked in Fig.\ref{map},   are given in Table \ref{colors}. The colours were corrected for extinction  $E_{B-V}=0.3$, estimated from $H_{\alpha}/H_{\beta}$ ratio. Note however that the assumption of uniform distribution of $E_{B-V}$ may be the source of some uncertainty of colour estimation. 
  \begin{table} 
  \caption{Colour indices corrected for extinction of the regions of NGC~90 shown in Fig. \ref{map}. }\label{colors}
  \begin{center}
   \begin{tabular}{lcc}
      \hline
   region&u-g&g-r\\
      &mag.&mag.\\
      \hline
   1&1.55$\pm$0.08&0.61$\pm$0.04\\
   2&1.16$\pm$0.18&0.45$\pm$0.10\\
   3&1.12$\pm$0.13&0.29$\pm$ 0.07\\
   4&1.03$\pm$0.13&0.26$\pm$0.07\\
   5&0.09$\pm$0.22&0.16$\pm$0.17\\
   6&0.25$\pm$0.30&0.10$\pm$0.23\\
   7&-0.15$\pm$0.26&-0.14$\pm$0.23\\
   8&0.26$\pm$0.26&-0.05$\pm$0.20\\
   9&0.36$\pm$0.27&0.11$\pm$0.20\\
   10&0.29$\pm$0.27&-0.07$\pm$0.21\\
   11&0.89$\pm$0.22&-0.01$\pm$0.14\\
   12&1.23$\pm$0.37&0.29$\pm$0.20\\
   \hline
\end{tabular}
  \end{center}
  \end{table}
    Fig. \ref{arp65_ug_gr} illustrates a position of these regions at two-colour diagram (the numbered symbols). We also show there  the model Starburst99 tracks \citep{Leitherer1999} for continuous (dashed line) and instantaneous (black line)  star formation (for Kroupa IMF \citealt{Kroupa2001}). For comparison, we also placed there the normal colour sequence (NCS) that galaxies of various morphological types form. To obtain this sequence we took the $U-B$ and $B-V$ colours of the galaxies of different morphological types brighter than $B_T = 11$ from Hyperleda database and converted these colours to $(u-g)$ and $(g-r)$   using the coefficients from \citet{Jester2005}. The band of NCS reflects the dependence between colour indices and the relative contribution of young stars to the total luminosity. It can also be considered as the sequence of colours for different time scales of star formation decay. The most red galaxies with a very low star formation rate occupy the upper part of the sequence. Regions of the outburst of star formation, or, conversely, of its rapid termination, should deviate  from NCS.

As the diagram shows,  all numbered regions (with the exception of region 11) can be divided into two groups. First, a center of  the galaxy (1) and the regions of bright spiral arms (2,3,4,12) lie along the NCS. They contain a mixture of stars of different ages, and by their stellar composition they are similar to spiral galaxies of Sa-Sc types, with a moderate rate of star formation. A colour of the central region (1) also evidences normal stellar population, consisting mainly of old stars. The second group of regions (5, 6, 7, 8, 9, 10) does not concord with NCS, being the areas of active birth of stars: their characteristic age is only a few million years. All these areas lie on the continuation of spiral arms outside the inner regular spiral pattern (except for the youngest region 7, located outside of the spirals), where tidal effects, responsible for the distortion of the spiral structure, influence the properties of interstellar medium. A current or recent star formation in the local areas can be associated with colliding gas flows inspired by tidal impact. 
   
   The extended region 11 of enhanced brightness is located in the tidal tail furthest from the main body of the galaxy. It should have a slightly different story: its color is well described by the model where all stars emerged at a time of  about $\sim$ 165 Myr ago, which may correspond to the time of the beginning of formation of tidal tails. This age is consistent with the estimate of \citet{Sengupta2015} which claim that the shortest possible time since the closest approach of galaxies to be $\sim1.9\times10^8$ years ago.

\subsection{On the nature of \HI cloud in NGC~90}

A unique feature of  NGC~90 is the anomalous velocity of gas in the vast region in SE part of the galaxy which may be interpreted as a giant non-rotating massive \HI cloud, strongly shifted from the centre of galaxy \citep{Sengupta2015}. It is not inferior in size to the optical galaxy, and its mass is approximately equal to the gas mass in the main disc of NGC~90 (about 3.4$\times10^9 M_\odot$). LOS velocity of the `cloud' is at 340 \kms   higher than  the velocity of a galaxy centre, and its maximum column density of \HI exceeds that observed in the main galaxy. The nature of this cloud remains unclear. The cited authors came to conclusion that the most reasonable explanation of the SE high-velocity \HI mass is the tidal interaction between NGC~90 and NGC~93 which caused a displacement of about half of the \HI beyond the main disc. As an example of similar event they refer to the interacting pair Arp 181 \citep{Sengupta2013}, and some other examples where \HI mass displacement is also observed. 

Note however that  tidal interaction {unavoidably}  involves a stellar disc population parallel with diffuse gas, so the \HI tidal displacement in interacting  galaxies is usually accompanied by the stellar flows. In our case there is no any observed trace of diffuse  low brightness stellar background in the \HI cloud. We also note that a velocity variation along tidal tails in interacting systems is usually smooth and have a small velocity gradient, so even  in a tip of a tail the velocity of gas does not strongly differ from that of parent galaxy. In NGC~90 we have quite different picture: according to \citet{Sengupta2015}, velocity distributions of the  \HI cloud and the  disc gas almost do not overlap, their maxima being  widely separated (see also a two-humped velocity distribution in the single dish spectrum of \HI obtained earlier for this galaxy by \citealt{Springob2005}). The observed velocity of the most swift blob of gas in the `cloud' exceeds the galaxy velocity more than at 400 \kms, while the velocity of disc rotation is twice slower,  which is hard to explain by the tidal interaction only. 

At last, a high velocity of gas `cloud' with respect to the galaxy gives evidence that the gas moves in a direction close to the line of sight. At the same time we see the optical tidal distortion of the galaxy disc, namely the optical tidal tails that continue spiral arms (also containing a gas), which are  stretched in the direction  far from LOS. 
{It is very unlikely that the gas tail shares the same origin as the tidal tails due to it being offset in a different plane. Therefore, a common origin of the optical tails and a gas stream observed as the \HI `cloud' looks incredible. 
}
Another  mechanism that can effectively remove a significant part of the gas from the galactic disc in a time  short compared with the time of gas consumption is the ram pressure of intergalactic medium  in a group or in a cluster where a galaxy moves. Indeed, galaxies in groups with
diffuse X-ray emission are often \HI deficient, and have lost more gas compared to those in groups without X-ray emission, which at least in part may be due to ram pressure \citep{Sengupta2006}. 

However, as \citet{Sengupta2015} rightly noted, one can  expect the ram pressure in groups to be at least an order of magnitude lower than in clusters.  Nevertheless it is evident that  the resultant gas loss depends on specific circumstances. For example, in Virgo cluster, where the effective gas sweeping is well established \citep[see e.g.][]{Chung2009, Safonova2011}  a mean gas density  of intracluster medium is not too high:  based on the ROSAT X-ray observations, electron volume density  $n_e$ in the central region  at about 0.2 Mpc from M87 was estimated to be near $10^{-3} cm^{-3}$ \citep{Nulsen1995}, and it falls down approximately inversely proportional to the radial distance $R_{cl}$. In the frame of $\beta$ model of gas distribution,  \citet{Boselli2016} estimated $n_e$ in Virgo cluster as low as $ ~10^{-5} cm^{-3}$ at $R_{cl} \approx 0.5$ Mpc in the region of spiral galaxy NGC~4569, which has undergone a ram pressure stripping event.  In big groups (or small clusters) of galaxies with the luminosity of hot gas $L_x \sim$  several $10^{42}$ erg/s an intra-group gas density is of the same order as in the central region of Virgo, that is about $10^{-3} cm^{-3}$ \citep{Dahlem2000,Sengupta2006}. Hence a difference of ram pressure $P\sim n_eV^2$ between groups and clusters determines primarily by a relative velocity of galaxies $V$, which in the mean  is several times lower in groups than in rich clusters.

Note that the luminosity $L_x$ of the  group to which the NGC~90 belongs, is rather high: according to ROSAT data,  $L_x\approx  3\times 10^{42}$ erg/s \citep{Mahdavi2004}, so the group is rich of hot gas.  In our case it is the most important that NGC~90 has unusually large LOS velocity  with respect to a systemic velocity of the group: according to \citet{Sengupta2015} this LOS velocity difference is 574 \kms, so the effect of ram pressure may be strong enough to deprive the galaxy of a significant part of gas located in the outer regions of the disc. 

An indirect evidence of the origin of the SE cloud as the result of ram pressure is the proximity of the `cloud' velocity { to the mean velocity of group members. The parent galaxy deviates more from the group by its velocity.} Noteworthy is also the high column density of gas in the `cloud', which exceeds the maximum gas density in the main galaxy, and the absence of noticeable regions of active star formation  in the  most dense region of the `cloud'. This clearly evidences, that the `cloud'  most probably is stretched along the line-of-sight, so the mean volume density of gas, unlike a column density, remains low. In this scenario a  high velocity of NGC~90 relative to the other  group members indicates that it's the first infall of the galaxy into the central region of the group, which led to its gas loss. It agrees with the  observed ''cloud''  offset away from the group center as { is} clearly evident from Fig.~2 in \citet{Sengupta2015} which shows the locations of Arp~65 and other group members together with  intra-group medium X-ray emission centroid.

The gas  will be removed from a stellar disc if the force of ram pressure of  intragroup gas (IGG) exceeds the restoring gravitational force per unit area \citep{GunnGott1972}, which leads to the simplified equation:
$$\rho_{icm}V_g^2 >  2\pi G\Sigma_{gas}\Sigma_*,$$
where $\rho_{icm}, V_g, \Sigma_{gas}, \Sigma_{*}$ are the IGG density, the velocity of the galaxy through the IGG perpendicular to the disc plane, mass surface densities of the gas disc and of the stellar disc, respectively. If, by analogy with the normal spiral galaxies, to assume that the interstellar  gas density beyond the inner region of a   disc is $\Sigma_{gas}<10M\odot/pc^2$ and to take $\rho_{ICM}$ for intra-group gas, corresponding to $n_e = 10^{-3}cm^{-3}$,  then  the resulting threshold stellar density $$ \Sigma_{*} \approx 25M_\odot/pc^2.$$ On the other hand, a SDSS-based photometry  of NGC~90 gives for radial distance corresponding to borderline of  the  inner symmetric spiral arms  a brightness  $\mu_g = 23.7 \mbox{mag}\, \mbox{arcsec}^{-2}$ and colour $(g-r) = 0.6$ mag (non-corrected for extinction). Then the colour - M/L relation \citep{Roediger2015} leads to the expected  surface density $\Sigma_{*} = 30\pm4M_\odot/pc^2$, which is close to the expected density  level corresponding to  gas removal.   Whilst total gas stripping will not occur, the inside-out nature of RPS means that the outer gas will likely  be swept out of the disc.  Hence a ram pressure  
could indeed deprive the galaxy of a significant part of  gas from the outer regions of its disc, when it enters the inner area of a group. Since the velocity of NGC~90 is lower than the mean velocity of the group,  the galaxy apparently lost its gas, entering the rear side of the group, moving  toward the observer.

A possible example of similar gas stripped tail demonstrates a jellyfish galaxy JO206, observed in \HI line by \citet{Ramatsoku2019}. It is located near the centre of the low-mass galaxy cluster IIZw108, having $L_X \sim 10^{44}\ \mathrm{erg\ s^{-1}}$. As in the case of NGC~90, JO206  appears to be a galaxy falling into the cluster, which experiences a ram pressure. A distribution of \HI in JO206 is perturbed and exhibits a one-sided, $\sim$ 90 kpc-long  \HI tail, emanating from the optical disc. Its hydrogen mass   (about  $3\times 10^9 M_\odot$)  is the same as in the ''cloud'' of NGC~90. A variation of LOC  velocity along the tail is also high, exceeding 300 \kms. There are  numerous star-forming regions scattered along the tail of JO206. In the case of NGC~90 we don't see any bright areas of active star formation connected with \HI cloud: perhaps we are observing this galaxy at the later stage of gas sweeping. Nonetheless a chain of low luminous \HII regions crossed by the slit PA = $125^\circ$ with LOS velocities laying between those for the galaxy and the cloud, are clearly seen in Fig.\ref{arp65_results} (right panel) evidencing that small local sites of star formation also exist in the ejected  gas flow.  

The another example of a massive stripped tail of neutral gas was found in a jellyfish galaxy JO204 in the low mass cluster  { Abell 957} \citep{Deb2020}. This 90-kpc long ram-pressure stripped tail  points away from the cluster center and contains most of \HI gas of the parent galaxy. The galaxy  also possesses  a more short and very clumpy ionized gas flows connecting with \HI tail,  where star formation remains active.  

Actually in the case of NGC~90 the picture is complicated by  gravitational interaction of galaxy with the neighbour galaxy NGC~93, a photometrically estimated mass of which, according to \citet{Sengupta2015}, is $1.4\times10^{11}M_\odot$. The difference of LOS between NGC~93 and NGC~90 exceeds 300 \kms, therefore, with high probability, galaxies are not bounded gravitationally, although a strong morphological distortion of NGC~90 evidences a tidal interaction. Tidal forces  could play a role in the shear and stretching of the gas flow,  weakening a connection between  gas expelled from the disc and the parent galaxy. However, the tidal stretching remains unnoticeable on the \HI maps, since, as the above arguments show, the gas flow from NGC~90 is apparently extended along LOS.

\section{Conclusion}

Spectral observations and the analysis of photometric data of NGC~90 confirmed that this is a moderate-mass galaxy with gas metallicity slightly less than the  solar one. It has the asymmetric distribution of emission regions  of current or recent star-formation outside of the regular spiral arms, and predominantly shock excitation of gas in the central part of the galaxy and at least in some regions of spiral arms. A nucleus of galaxy possess a LINER-like activity.

The main peculiarity of NGC~90 is the presence of massive `cloud' of \HI projecting onto its SE edge \citep{Sengupta2015}. The cloud LOS velocity is significantly  different from the velocity of the adjacent part of the galaxy, and therefore it is loosely  connected with the latter. Observations showed the presence of emission regions in the high-velocity gas, apparently related to local sites of star formation triggered by interaction. 

We argue that the galaxy was subjected to two types of external influence. Firstly, this is a tidal interaction with the neighbour galaxy NGC~93. Apparently  the convergence of galaxies took place 150-200 Myr ago. It manifests itself not only in the morphology of a spiral galaxy, but also in a change in the outer disc orientation, which, unlike the inner disc, is observed close to face-on, and also in a low radial gradient of gas metallicity, as well as in presence of tidal tails.

 Secondly, it is a ram pressure of the intra-group hot gas onto the disc, which led to a loss of a significant part of the diffuse  gas of the galaxy. The gaseous tail, which inevitably arises in such cases, is apparently elongated along the line of sight, which creates the illusion of a cloud with a high column density of gas projecting onto the SE-part of the galaxy. A chain of local emission regions in this tail evidences that star formation inspired by the high gas pressure, has not yet completely faded.

\section*{Acknowledgements} 

This study was supported by Russian Foundation for Basic Research (project 18-32-20120). AS research was supported by The Russian Science Foundation (RSCF) grant No. 19-12-00281. AZ, AS and OE are grateful the Program of development of M.V. Lomonosov Moscow State University (Leading Scientific School `Physics of stars, relativistic objects and galaxies') for the financial support. O.E. acknowledges the support from Foundation of development of theoretical physics and mathematics `Basis'.
Observations conducted with the 6-m telescope of the Special Astrophysical Observatory of the Russian Academy of Sciences carried out with the financial support of the Ministry of Science and Higher Education of the Russian Federation (including agreement No. 05.619.21.0016, project ID RFMEFI61919X0016). {We thank the anonymous referee for comments that helped us to improve the paper}.

In this work we used the archival data of SDSS-IV and DECaLS surveys. 
SDSS-IV is managed by the Astrophysical Research Consortium for the 
Participating Institutions of the SDSS Collaboration including the 
Brazilian Participation Group, the Carnegie Institution for Science, 
Carnegie Mellon University, the Chilean Participation Group, the French Participation Group, Harvard-Smithsonian Center for Astrophysics, 
Instituto de Astrof\'isica de Canarias, The Johns Hopkins University, 
Kavli Institute for the Physics and Mathematics of the Universe (IPMU) / 
University of Tokyo, Lawrence Berkeley National Laboratory, 
Leibniz Institut f\"ur Astrophysik Potsdam (AIP),  
Max-Planck-Institut f\"ur Astronomie (MPIA Heidelberg), 
Max-Planck-Institut f\"ur Astrophysik (MPA Garching), 
Max-Planck-Institut f\"ur Extraterrestrische Physik (MPE), 
National Astronomical Observatories of China, New Mexico State University, 
New York University, University of Notre Dame, 
Observat\'ario Nacional / MCTI, The Ohio State University, 
Pennsylvania State University, Shanghai Astronomical Observatory, 
United Kingdom Participation Group,
Universidad Nacional Aut\'onoma de M\'exico, University of Arizona, 
University of Colorado Boulder, University of Oxford, University of Portsmouth, 
University of Utah, University of Virginia, University of Washington, University of Wisconsin, 
Vanderbilt University, and Yale University.
 
The Legacy Surveys consist of three individual and complementary projects: the Dark Energy Camera Legacy Survey (DECaLS; NOAO Proposal ID  2014B-0404; PIs: David Schlegel and Arjun Dey), the Beijing-Arizona Sky Survey (BASS; NOAO Proposal ID 2015A-0801; PIs: Zhou Xu and Xiaohui Fan), and the Mayall z-band Legacy Survey (MzLS; NOAO Proposal ID  2016A-0453; PI: Arjun Dey). DECaLS, BASS and MzLS together include data obtained, respectively, at the Blanco telescope, Cerro Tololo Inter-American Observatory, National Optical Astronomy Observatory (NOAO); the Bok telescope, Steward Observatory, University of Arizona; and the Mayall telescope, Kitt Peak National Observatory, NOAO. The Legacy Surveys project is honored to be permitted to conduct astronomical research on Iolkam Du'ag (Kitt Peak), a mountain with particular significance to the Tohono O'odham Nation.

  \label{lastpage}


\begin{thebibliography}{}
	\makeatletter
	\relax
	\def\mn@urlcharsother{\let\do\@makeother \do\$\do\&\do\#\do\^\do\_\do\%\do\~}
	\def\mn@doi{\begingroup\mn@urlcharsother \@ifnextchar [ {\mn@doi@}
		{\mn@doi@[]}}
	\def\mn@doi@[#1]#2{\def\@tempa{#1}\ifx\@tempa\@empty \href
		{http://dx.doi.org/#2} {doi:#2}\else \href {http://dx.doi.org/#2} {#1}\fi
		\endgroup}
	\def\mn@eprint#1#2{\mn@eprint@#1:#2::\@nil}
	\def\mn@eprint@arXiv#1{\href {http://arxiv.org/abs/#1} {{\tt arXiv:#1}}}
	\def\mn@eprint@dblp#1{\href {http://dblp.uni-trier.de/rec/bibtex/#1.xml}
		{dblp:#1}}
	\def\mn@eprint@#1:#2:#3:#4\@nil{\def\@tempa {#1}\def\@tempb {#2}\def\@tempc
		{#3}\ifx \@tempc \@empty \let \@tempc \@tempb \let \@tempb \@tempa \fi \ifx
		\@tempb \@empty \def\@tempb {arXiv}\fi \@ifundefined
		{mn@eprint@\@tempb}{\@tempb:\@tempc}{\expandafter \expandafter \csname
			mn@eprint@\@tempb\endcsname \expandafter{\@tempc}}}
	
	\bibitem[\protect\citeauthoryear{{Afanasiev} \& {Moiseev}}{{Afanasiev} \&
		{Moiseev}}{2011}]{AfanasievMoiseev2011}
	{Afanasiev} V.~L.,  {Moiseev} A.~V.,  2011, Baltic Astronomy, \href
	{http://adsabs.harvard.edu/abs/2011BaltA..20..363A} {20, 363}
	
	\bibitem[\protect\citeauthoryear{Allen, Groves, Dopita, Sutherland  \&
		Kewley}{Allen et~al.}{2008}]{Allen2008}
	Allen M.~G.,  Groves B.~A.,  Dopita M.~A.,  Sutherland R.~S.,   Kewley L.~J.,
	2008, \apjs, 178, 20
	
	\bibitem[\protect\citeauthoryear{{Baldwin}, {Phillips}  \&
		{Terlevich}}{{Baldwin} et~al.}{1981}]{BPT}
	{Baldwin} J.~A.,  {Phillips} M.~M.,   {Terlevich} R.,  1981, \mn@doi [\pasp]
	{10.1086/130766}, \href {http://adsabs.harvard.edu/abs/1981PASP...93....5B}
	{93, 5}
	
	\bibitem[\protect\citeauthoryear{{Boselli} et~al.,}{{Boselli}
		et~al.}{2016}]{Boselli2016}
	{Boselli} A.,  et~al., 2016, \mn@doi [\aap] {10.1051/0004-6361/201527795},
	\href {https://ui.adsabs.harvard.edu/abs/2016A&A...587A..68B} {587, A68}
	
	\bibitem[\protect\citeauthoryear{{Cardelli}, {Clayton}  \& {Mathis}}{{Cardelli}
		et~al.}{1989}]{Cardelli1989}
	{Cardelli} J.~A.,  {Clayton} G.~C.,   {Mathis} J.~S.,  1989, \mn@doi [\apj]
	{10.1086/167900}, \href {http://adsabs.harvard.edu/abs/1989ApJ...345..245C}
	{345, 245}
	
	\bibitem[\protect\citeauthoryear{{Casteels} et~al.,}{{Casteels}
		et~al.}{2013}]{Casteels2013}
	{Casteels} K. R.~V.,  et~al., 2013, \mn@doi [\mnras] {10.1093/mnras/sts391},
	\href {https://ui.adsabs.harvard.edu/abs/2013MNRAS.429.1051C} {429, 1051}
	
	\bibitem[\protect\citeauthoryear{{Chilingarian}, {Prugniel}, {Sil'Chenko}  \&
		{Koleva}}{{Chilingarian} et~al.}{2007a}]{Chilingarian2007a}
	{Chilingarian} I.,  {Prugniel} P.,  {Sil'Chenko} O.,   {Koleva} M.,  2007a, in
	{Vazdekis} A.,  {Peletier} R.,  eds,  IAU Symposium Vol. 241, IAU Symposium.
	pp 175--176 (\mn@eprint {arXiv} {0709.3047}),
	\mn@doi{10.1017/S1743921307007752}
	
	\bibitem[\protect\citeauthoryear{{Chilingarian}, {Prugniel}, {Sil'Chenko}  \&
		{Afanasiev}}{{Chilingarian} et~al.}{2007b}]{Chilingarian2007b}
	{Chilingarian} I.~V.,  {Prugniel} P.,  {Sil'Chenko} O.~K.,   {Afanasiev} V.~L.,
	2007b, \mn@doi [\mnras] {10.1111/j.1365-2966.2007.11549.x}, \href
	{http://adsabs.harvard.edu/abs/2007MNRAS.376.1033C} {376, 1033}
	
	\bibitem[\protect\citeauthoryear{{Chung}, {van Gorkom}, {Kenney}, {Crowl}  \&
		{Vollmer}}{{Chung} et~al.}{2009}]{Chung2009}
	{Chung} A.,  {van Gorkom} J.~H.,  {Kenney} J. D.~P.,  {Crowl} H.,   {Vollmer}
	B.,  2009, \mn@doi [\aj] {10.1088/0004-6256/138/6/1741}, \href
	{https://ui.adsabs.harvard.edu/abs/2009AJ....138.1741C} {138, 1741}
	
	\bibitem[\protect\citeauthoryear{{Dahlem} \& {Thiering}}{{Dahlem} \&
		{Thiering}}{2000}]{Dahlem2000}
	{Dahlem} M.,  {Thiering} I.,  2000, \mn@doi [\pasp] {10.1086/316511}, \href
	{https://ui.adsabs.harvard.edu/abs/2000PASP..112..148D} {112, 148}
	
	\bibitem[\protect\citeauthoryear{{Deb} et~al.,}{{Deb} et~al.}{2020}]{Deb2020}
	{Deb} T.,  et~al., 2020, \mn@doi [\mnras] {10.1093/mnras/staa968}, \href
	{https://ui.adsabs.harvard.edu/abs/2020MNRAS.494.5029D} {494, 5029}
	
	\bibitem[\protect\citeauthoryear{{Dopita} et~al.,}{{Dopita}
		et~al.}{2005}]{Dopita2005}
	{Dopita} M.~A.,  et~al., 2005, \mn@doi [\apj] {10.1086/423948}, \href
	{https://ui.adsabs.harvard.edu/abs/2005ApJ...619..755D} {619, 755}
	
	\bibitem[\protect\citeauthoryear{{Fitzpatrick}}{{Fitzpatrick}}{1999}]{Fitzpatrick1999}
	{Fitzpatrick} E.~L.,  1999, \mn@doi [\pasp] {10.1086/316293}, \href
	{http://adsabs.harvard.edu/abs/1999PASP..111...63F} {111, 63}
	
	\bibitem[\protect\citeauthoryear{{Gunn} \& {Gott}}{{Gunn} \&
		{Gott}}{1972}]{GunnGott1972}
	{Gunn} J.~E.,  {Gott} J.~Richard I.,  1972, \mn@doi [\apj] {10.1086/151605},
	\href {https://ui.adsabs.harvard.edu/abs/1972ApJ...176....1G} {176, 1}
	
	\bibitem[\protect\citeauthoryear{{Haffner} et~al.,}{{Haffner}
		et~al.}{2009}]{Haffner2009}
	{Haffner} L.~M.,  et~al., 2009, \mn@doi [Reviews of Modern Physics]
	{10.1103/RevModPhys.81.969}, \href
	{https://ui.adsabs.harvard.edu/abs/2009RvMP...81..969H} {81, 969}
	
	\bibitem[\protect\citeauthoryear{{Ho}, {Kudritzki}, {Kewley}, {Zahid},
		{Dopita}, {Bresolin}  \& {Rupke}}{{Ho} et~al.}{2015}]{Ho2015}
	{Ho} I.-T.,  {Kudritzki} R.-P.,  {Kewley} L.~J.,  {Zahid} H.~J.,  {Dopita}
	M.~A.,  {Bresolin} F.,   {Rupke} D.~S.~N.,  2015, \mn@doi [\mnras]
	{10.1093/mnras/stv067}, \href
	{http://adsabs.harvard.edu/abs/2015MNRAS.448.2030H} {448, 2030}
	
	\bibitem[\protect\citeauthoryear{{Jedrzejewski}}{{Jedrzejewski}}{1987}]{Jedrzejewski1987}
	{Jedrzejewski} R.~I.,  1987, \mn@doi [\mnras] {10.1093/mnras/226.4.747}, \href
	{http://adsabs.harvard.edu/abs/1987MNRAS.226..747J} {226, 747}
	
	\bibitem[\protect\citeauthoryear{{Jester} et~al.,}{{Jester}
		et~al.}{2005}]{Jester2005}
	{Jester} S.,  et~al., 2005, \mn@doi [\aj] {10.1086/432466}, \href
	{https://ui.adsabs.harvard.edu/abs/2005AJ....130..873J} {130, 873}
	
	\bibitem[\protect\citeauthoryear{{Kauffmann} et~al.,}{{Kauffmann}
		et~al.}{2003}]{Kauffmann03}
	{Kauffmann} G.,  et~al., 2003, \mn@doi [\mnras]
	{10.1111/j.1365-2966.2003.07154.x}, \href
	{http://adsabs.harvard.edu/abs/2003MNRAS.346.1055K} {346, 1055}
	
	\bibitem[\protect\citeauthoryear{{Kewley}, {Dopita}, {Sutherland}, {Heisler}
		\& {Trevena}}{{Kewley} et~al.}{2001}]{Kewley2001}
	{Kewley} L.~J.,  {Dopita} M.~A.,  {Sutherland} R.~S.,  {Heisler} C.~A.,
	{Trevena} J.,  2001, \mn@doi [\apj] {10.1086/321545}, \href
	{http://adsabs.harvard.edu/abs/2001ApJ...556..121K} {556, 121}
	
	\bibitem[\protect\citeauthoryear{{Kewley}, {Geller}  \& {Barton}}{{Kewley}
		et~al.}{2006}]{Kewley2006}
	{Kewley} L.~J.,  {Geller} M.~J.,   {Barton} E.~J.,  2006, \mn@doi [\aj]
	{10.1086/500295}, \href {http://adsabs.harvard.edu/abs/2006AJ....131.2004K}
	{131, 2004}
	
	\bibitem[\protect\citeauthoryear{{Kewley}, {Rupke}, {Zahid}, {Geller}  \&
		{Barton}}{{Kewley} et~al.}{2010}]{Kewley2010}
	{Kewley} L.~J.,  {Rupke} D.,  {Zahid} H.~J.,  {Geller} M.~J.,   {Barton} E.~J.,
	2010, \mn@doi [\apjl] {10.1088/2041-8205/721/1/L48}, \href
	{http://adsabs.harvard.edu/abs/2010ApJ...721L..48K} {721, L48}
	
	\bibitem[\protect\citeauthoryear{{Kroupa}}{{Kroupa}}{2001}]{Kroupa2001}
	{Kroupa} P.,  2001, \mn@doi [\mnras] {10.1046/j.1365-8711.2001.04022.x}, \href
	{http://adsabs.harvard.edu/abs/2001MNRAS.322..231K} {322, 231}
	
	\bibitem[\protect\citeauthoryear{{Kumari}, {Maiolino}, {Belfiore}  \&
		{Curti}}{{Kumari} et~al.}{2019}]{Kumari2019}
	{Kumari} N.,  {Maiolino} R.,  {Belfiore} F.,   {Curti} M.,  2019, \mn@doi
	[\mnras] {10.1093/mnras/stz366}, \href
	{http://adsabs.harvard.edu/abs/2019MNRAS.485..367K} {485, 367}
	
	\bibitem[\protect\citeauthoryear{{Lacerda} et~al.,}{{Lacerda}
		et~al.}{2018}]{Lacerda2018}
	{Lacerda} E.~A.~D.,  et~al., 2018, \mn@doi [\mnras] {10.1093/mnras/stx3022},
	\href {http://adsabs.harvard.edu/abs/2018MNRAS.474.3727L} {474, 3727}
	
	\bibitem[\protect\citeauthoryear{{Le Borgne}, {Rocca-Volmerange}, {Prugniel},
		{Lan{\c c}on}, {Fioc}  \& {Soubiran}}{{Le Borgne}
		et~al.}{2004}]{LeBorgneetal2004}
	{Le Borgne} D.,  {Rocca-Volmerange} B.,  {Prugniel} P.,  {Lan{\c c}on} A.,
	{Fioc} M.,   {Soubiran} C.,  2004, \mn@doi [\aap]
	{10.1051/0004-6361:200400044}, \href
	{http://adsabs.harvard.edu/abs/2004A%26A...425..881L} {425, 881}
		
		\bibitem[\protect\citeauthoryear{{Leitherer} et~al.,}{{Leitherer}
			et~al.}{1999}]{Leitherer1999}
		{Leitherer} C.,  et~al., 1999, \mn@doi [\apjs] {10.1086/313233}, \href
		{http://adsabs.harvard.edu/abs/1999ApJS..123....3L} {123, 3}
		
		\bibitem[\protect\citeauthoryear{{Mahdavi} \& {Geller}}{{Mahdavi} \&
			{Geller}}{2004}]{Mahdavi2004}
		{Mahdavi} A.,  {Geller} M.~J.,  2004, \mn@doi [\apj] {10.1086/383458}, \href
		{https://ui.adsabs.harvard.edu/abs/2004ApJ...607..202M} {607, 202}
		
		\bibitem[\protect\citeauthoryear{{Makarov}, {Prugniel}, {Terekhova}, {Courtois}
			\& {Vauglin}}{{Makarov} et~al.}{2014}]{Makarov2014}
		{Makarov} D.,  {Prugniel} P.,  {Terekhova} N.,  {Courtois} H.,   {Vauglin} I.,
		2014, \mn@doi [\aap] {10.1051/0004-6361/201423496}, \href
		{http://adsabs.harvard.edu/abs/2014A%26A...570A..13M} {570, A13}
			
			\bibitem[\protect\citeauthoryear{{Marino} et~al.,}{{Marino}
				et~al.}{2013}]{Marino2013}
			{Marino} R.~A.,  et~al., 2013, \mn@doi [\aap] {10.1051/0004-6361/201321956},
			\href {http://adsabs.harvard.edu/abs/2013A%26A...559A.114M} {559, A114}
				
				\bibitem[\protect\citeauthoryear{{Nulsen} \& {Bohringer}}{{Nulsen} \&
					{Bohringer}}{1995}]{Nulsen1995}
				{Nulsen} P.~E.~J.,  {Bohringer} H.,  1995, \mn@doi [\mnras]
				{10.1093/mnras/274.4.1093}, \href
				{https://ui.adsabs.harvard.edu/abs/1995MNRAS.274.1093N} {274, 1093}
				
				\bibitem[\protect\citeauthoryear{{Oh}, {Kim}, {Lee}  \& {Kim}}{{Oh}
					et~al.}{2008}]{Oh2008}
				{Oh} S.~H.,  {Kim} W.-T.,  {Lee} H.~M.,   {Kim} J.,  2008, \mn@doi [\apj]
				{10.1086/588184}, \href
				{https://ui.adsabs.harvard.edu/abs/2008ApJ...683...94O} {683, 94}
				
				\bibitem[\protect\citeauthoryear{{Osterbrock} \& {Ferland}}{{Osterbrock} \&
					{Ferland}}{2006}]{Osterbrock2006}
				{Osterbrock} D.~E.,  {Ferland} G.~J.,  2006, {Astrophysics of gaseous nebulae
					and active galactic nuclei}, 2nd edn.
				University Science Books
				
				\bibitem[\protect\citeauthoryear{{Pilyugin} \& {Grebel}}{{Pilyugin} \&
					{Grebel}}{2016}]{Pilyugin16}
				{Pilyugin} L.~S.,  {Grebel} E.~K.,  2016, \mn@doi [\mnras]
				{10.1093/mnras/stw238}, \href
				{http://adsabs.harvard.edu/abs/2016MNRAS.457.3678P} {457, 3678}
				
				\bibitem[\protect\citeauthoryear{{Pilyugin}, {V{\'\i}lchez}  \&
					{Contini}}{{Pilyugin} et~al.}{2004}]{Pilyugin2004}
				{Pilyugin} L.~S.,  {V{\'\i}lchez} J.~M.,   {Contini} T.,  2004, \mn@doi [\aap]
				{10.1051/0004-6361:20034522}, \href
				{https://ui.adsabs.harvard.edu/abs/2004A&A...425..849P} {425, 849}
				
				\bibitem[\protect\citeauthoryear{{Pilyugin}, {Grebel}  \& {Kniazev}}{{Pilyugin}
					et~al.}{2014}]{Pilyugin2014}
				{Pilyugin} L.~S.,  {Grebel} E.~K.,   {Kniazev} A.~Y.,  2014, \mn@doi [\aj]
				{10.1088/0004-6256/147/6/131}, \href
				{http://adsabs.harvard.edu/abs/2014AJ....147..131P} {147, 131}
				
				\bibitem[\protect\citeauthoryear{{Ponomareva}, {Verheijen}, {Peletier}  \&
					{Bosma}}{{Ponomareva} et~al.}{2017}]{Ponomareva2017}
				{Ponomareva} A.~A.,  {Verheijen} M. A.~W.,  {Peletier} R.~F.,   {Bosma} A.,
				2017, \mn@doi [\mnras] {10.1093/mnras/stx1018}, \href
				{https://ui.adsabs.harvard.edu/abs/2017MNRAS.469.2387P} {469, 2387}
				
				\bibitem[\protect\citeauthoryear{{Ramatsoku} et~al.,}{{Ramatsoku}
					et~al.}{2019}]{Ramatsoku2019}
				{Ramatsoku} M.,  et~al., 2019, \mn@doi [\mnras] {10.1093/mnras/stz1609}, \href
				{https://ui.adsabs.harvard.edu/abs/2019MNRAS.487.4580R} {487, 4580}
				
				\bibitem[\protect\citeauthoryear{{Rich}, {Torrey}, {Kewley}, {Dopita}  \&
					{Rupke}}{{Rich} et~al.}{2012}]{Rich2012}
				{Rich} J.~A.,  {Torrey} P.,  {Kewley} L.~J.,  {Dopita} M.~A.,   {Rupke}
				D.~S.~N.,  2012, \mn@doi [\apj] {10.1088/0004-637X/753/1/5}, \href
				{http://adsabs.harvard.edu/abs/2012ApJ...753....5R} {753, 5}
				
				\bibitem[\protect\citeauthoryear{{Roediger} \& {Courteau}}{{Roediger} \&
					{Courteau}}{2015}]{Roediger2015}
				{Roediger} J.~C.,  {Courteau} S.,  2015, \mn@doi [\mnras]
				{10.1093/mnras/stv1499}, \href
				{http://adsabs.harvard.edu/abs/2015MNRAS.452.3209R} {452, 3209}
				
				\bibitem[\protect\citeauthoryear{{Rosa}, {Dors}, {Krabbe}, {H{\"a}gele},
					{Cardaci}, {Pastoriza}, {Rodrigues}  \& {Winge}}{{Rosa}
					et~al.}{2014}]{Rosa2014}
				{Rosa} D.~A.,  {Dors} O.~L.,  {Krabbe} A.~C.,  {H{\"a}gele} G.~F.,  {Cardaci}
				M.~V.,  {Pastoriza} M.~G.,  {Rodrigues} I.,   {Winge} C.,  2014, \mn@doi
				[\mnras] {10.1093/mnras/stu1578}, \href
				{http://adsabs.harvard.edu/abs/2014MNRAS.444.2005R} {444, 2005}
				
				\bibitem[\protect\citeauthoryear{{Rupke}, {Kewley}  \& {Chien}}{{Rupke}
					et~al.}{2010}]{Rupke2010}
				{Rupke} D.~S.~N.,  {Kewley} L.~J.,   {Chien} L.-H.,  2010, \mn@doi [\apj]
				{10.1088/0004-637X/723/2/1255}, \href
				{http://adsabs.harvard.edu/abs/2010ApJ...723.1255R} {723, 1255}
				
				\bibitem[\protect\citeauthoryear{{Safonova}}{{Safonova}}{2011}]{Safonova2011}
				{Safonova} E.~S.,  2011, \mn@doi [Astronomy Reports]
				{10.1134/S1063772911110060}, \href
				{https://ui.adsabs.harvard.edu/abs/2011ARep...55.1016S} {55, 1016}
				
				\bibitem[\protect\citeauthoryear{{Sengupta} \& {Balasubramanyam}}{{Sengupta} \&
					{Balasubramanyam}}{2006}]{Sengupta2006}
				{Sengupta} C.,  {Balasubramanyam} R.,  2006, \mn@doi [\mnras]
				{10.1111/j.1365-2966.2006.10307.x}, \href
				{https://ui.adsabs.harvard.edu/abs/2006MNRAS.369..360S} {369, 360}
				
				\bibitem[\protect\citeauthoryear{{Sengupta}, {Dwarakanath}, {Saikia}  \&
					{Scott}}{{Sengupta} et~al.}{2013}]{Sengupta2013}
				{Sengupta} C.,  {Dwarakanath} K.~S.,  {Saikia} D.~J.,   {Scott} T.~C.,  2013,
				\mn@doi [\mnras] {10.1093/mnrasl/sls039}, \href
				{https://ui.adsabs.harvard.edu/abs/2013MNRAS.431L...1S} {431, L1}
				
				\bibitem[\protect\citeauthoryear{{Sengupta}, {Scott}, {Paudel}, {Saikia},
					{Dwarakanath}  \& {Sohn}}{{Sengupta} et~al.}{2015}]{Sengupta2015}
				{Sengupta} C.,  {Scott} T.~C.,  {Paudel} S.,  {Saikia} D.~J.,  {Dwarakanath}
				K.~S.,   {Sohn} B.~W.,  2015, \mn@doi [\aap] {10.1051/0004-6361/201425149},
				584, A114
				
				\bibitem[\protect\citeauthoryear{{Springob}, {Haynes}, {Giovanelli}  \&
					{Kent}}{{Springob} et~al.}{2005}]{Springob2005}
				{Springob} C.~M.,  {Haynes} M.~P.,  {Giovanelli} R.,   {Kent} B.~R.,  2005,
				\mn@doi [\apjs] {10.1086/431550}, \href
				{https://ui.adsabs.harvard.edu/abs/2005ApJS..160..149S} {160, 149}
				
				\bibitem[\protect\citeauthoryear{{Tody}}{{Tody}}{1986}]{iraf}
				{Tody} D.,  1986, in {Crawford} D.~L.,  ed.,  Society of Photo-Optical
				Instrumentation Engineers (SPIE) Conference Series Vol. 627, Instrumentation
				in astronomy VI. p.~733
				
				\bibitem[\protect\citeauthoryear{{Veilleux} \& {Osterbrock}}{{Veilleux} \&
					{Osterbrock}}{1987}]{BPT_S}
				{Veilleux} S.,  {Osterbrock} D.~E.,  1987, \mn@doi [\apjs] {10.1086/191166},
				\href {http://adsabs.harvard.edu/abs/1987ApJS...63..295V} {63, 295}
				
				\bibitem[\protect\citeauthoryear{{Zasov}, {Saburova}, {Katkov}, {Egorov}  \&
					{Afanasiev}}{{Zasov} et~al.}{2015}]{zasovetal2015}
				{Zasov} A.,  {Saburova} A.,  {Katkov} I.,  {Egorov} O.,   {Afanasiev} V.,
				2015, \mn@doi [\mnras] {10.1093/mnras/stv454}, \href
				{http://adsabs.harvard.edu/abs/2015MNRAS.449.1605Z} {449, 1605}
				
				\bibitem[\protect\citeauthoryear{{Zasov}, {Saburova}, {Egorov}  \&
					{Afanasiev}}{{Zasov} et~al.}{2016}]{zasovetal2016}
				{Zasov} A.~V.,  {Saburova} A.~S.,  {Egorov} O.~V.,   {Afanasiev} V.~L.,  2016,
				\mn@doi [\mnras] {10.1093/mnras/stw1905}, \href
				{http://adsabs.harvard.edu/abs/2016MNRAS.462.3419Z} {462, 3419}
				
				\bibitem[\protect\citeauthoryear{{Zasov}, {Saburova}, {Egorov}  \&
					{Uklein}}{{Zasov} et~al.}{2017}]{Zasovetal2017}
				{Zasov} A.~V.,  {Saburova} A.~S.,  {Egorov} O.~V.,   {Uklein} R.~I.,  2017,
				\mn@doi [\mnras] {10.1093/mnras/stx1158}, \href
				{http://adsabs.harvard.edu/abs/2017MNRAS.469.4370Z} {469, 4370}
				
				\bibitem[\protect\citeauthoryear{{Zasov}, {Saburova}, {Egorov}  \&
					{Afanasiev}}{{Zasov} et~al.}{2018}]{Zasovetal2018}
				{Zasov} A.~V.,  {Saburova} A.~S.,  {Egorov} O.~V.,   {Afanasiev} V.~L.,  2018,
				\mn@doi [\mnras] {10.1093/mnras/sty1017}, \href
				{http://adsabs.harvard.edu/abs/2018MNRAS.477.4908Z} {477, 4908}
				
				\bibitem[\protect\citeauthoryear{{Zasov}, {Saburova}, {Egorov}  \&
					{Dodonov}}{{Zasov} et~al.}{2019}]{Zasov2019}
				{Zasov} A.~V.,  {Saburova} A.~S.,  {Egorov} O.~V.,   {Dodonov} S.~N.,  2019,
				\mn@doi [\mnras] {10.1093/mnras/stz1025}, \href
				{https://ui.adsabs.harvard.edu/abs/2019MNRAS.486.2604Z} {486, 2604}
				
				\bibitem[\protect\citeauthoryear{{Zhang} et~al.,}{{Zhang}
					et~al.}{2017}]{Zhang2017}
				{Zhang} K.,  et~al., 2017, \mn@doi [\mnras] {10.1093/mnras/stw3308}, \href
				{https://ui.adsabs.harvard.edu/#abs/2017MNRAS.466.3217Z} {466, 3217}
				
				\bibitem[\protect\citeauthoryear{{van der Marel} \& {Franx}}{{van der Marel} \&
					{Franx}}{1993}]{vanderMarel1993}
				{van der Marel} R.~P.,  {Franx} M.,  1993, \mn@doi [\apj] {10.1086/172534},
				\href {http://adsabs.harvard.edu/abs/1993ApJ...407..525V} {407, 525}
				
				\makeatother
\end{thebibliography}
\end{document}